       \documentstyle [12pt]   {article}
 \evensidemargin\oddsidemargin
\begin{document}
\count0 = 1
 \title{  Quantum Information and    
 \\ Wave Function Collapse }
\author{ S.N.Mayburov \\
Lebedev Inst. of Physics\\
Leninsky Prospect 53, Moscow, Russia, 117924\\
E-mail :\quad   mayburov@sci.lebedev.ru}
\date {}
\maketitle

\begin{abstract}

  Information-theoretical restrictions on the
 information transferred in quantum measurements
are  regarded for
 the measurement of object $S$ performed via its interaction with 
   information system $O$. 
This information restrictions,  induced by Heisenberg commutation relations,
  are  derived in the formalism of inference maps in
Hilbert space. 
 $O$ restricted states $\xi^O$ are calculated from
  $S$,$O$ quantum dynamics
 and the structure of $O$ observables set (algebra).
%
  It's shown that this principal constraints on the information transfer
  result in
the appearance of stochasticity in the measurement outcomes.  Consequently,
   $\xi^O$ describes the  random 'pointer'
outcomes $q_j$,
which correspond to the collapse of $S$ pure state.
 $O$ decoherence by its environment is studied 
 for some $S$, $O$ systems, it's shown that it doesn't change
this results principally.
\end{abstract}


\section  {Introduction}

Despite its significant achievements, Quantum mechanics
 (QM) still contains some open questions concerned with its internal
 consistency$\cite {B1,Es2,J3}$. The most famous and oldest  of them is
  the State Collapse or Quantum Measurement
   problem$\cite {Em4,B5}$.
In this paper this problem 
  will be considered mainly 
within the framework of Information Theory $\cite {M6,S7,M8}$.
Really,   the  measurement of any system $S$ 
includes   the  transfer of information from
 $S$    to the information system  $O$  (Observer) which processes
 and memorizes it. Correspondingly, in Information Theory,
 any measuring system (MS)
can be described as the information channel, which transfers
the information about $S$ state to $O$ $\cite {M6}$.
%
Plainly, if some  restrictions on the information
 transfer via such channel exist,  they  can influence, in principle,
  the outcomes of measurement events percepted by $O$.
Recently,  it was shown that such
  constraints, induced mainly by Heisenberg commutation relations,
 result in the significant information losses for typical information
channels $\cite {HHH}$.
Our calculations for some simple MS models evidence that
 for MS such restrictions
  induce the unavoidable stochasticity in the measurements
of $S$ pure states, due to the loss of information about the rate of their
 purity \cite {M8}. Here we develop our formalism and present some new
results and their discussion.

For MS dynamics
 we shall exploit here the standard approach of measurement theory in which
  the evolution of all physical  objects,
 including $O$,
 is described  by the quantum state (density matrix) $\rho(t)$
 which  obeys 
 Schroedinger-Lioville equation in arbitrary reference frame (RF). 
The information-theoretical formalism of
 systems' self-description or 'measurement from inside' is applied
 to the  description of information acquisition by $O$  in
quantum measurements$\cite {S7}$.
 In  Schroedinger QM framework, it  realized by means of
 the formalism of inference maps in Hilbert space $\cite {B5}$.
 Overall, 
 we shall demonstrate for simple MS model  that such approach  
allows to construct the consistent measurement theory, which results in
the stochasticity of measurement outcomes  without exploit of additional
 Reduction Postulate in Schroedinger QM scheme.
It's instructive to note beforehand that even if  $O$ is the human brain,
in our theory the observer's consciousness doesn't play any role and
will not be referred to at all; some  aspects of this problem  
will be discussed  in conclusion.
 Yet for the illustrative purposes  some terms characteristic for
 conscious perception of signals will be used in the discussion of our model.


\section                   {Model of Quantum Measurements}

Here our measurement model will be described and some  aspects of
QM measurements theory  will be  reviewed. We shall also formulate 
the main ideas of our approach in semi-qualitative form,
 the detailed mathematical
 formalism will be described in the next chapter. 
 In our model MS consists of
  the studied system $S$,  detector $D$ and the information system
   $O$, which acquires and processes the incoming information.
The effect of MS decoherence by the environment$\cite {Z10}$
 will be considered in the final part of our paper, it will be shown
that in our theory its influence is inessential for the measurement process
  and so it will be analyzed separately.  
%
   $S$ is taken to be the particle with the spin $\frac {1}{2}$ and
 the measurement of
its projection  $S_z$  will be studied. Its $u,d$ eigenstates
 denoted $|s_{1,2}\rangle$, so that
  the  measured $S$ pure state is: 
$$
 \psi_s =a_1|s_1\rangle+a_2|s_2\rangle
$$
 For the comparison, the incoming
  $u,\, d$ 'test' mixture with the same $\bar{S}_z$ should be regarded 
also. This is $S$  ensemble described by the gemenge
$W^s=\{|s_i\rangle, \rm{P}_i \}$, where $\rm{P}_i$ $=|a_i|^2$
 are the probabilities
of  $|s_i\rangle$ in this ensemble \cite {B1},
 its  density matrix denoted $\rho^s_m$.
Analogously to $S$, in our model $D$ state in $O$ RF is 
 described by
Dirac vector $|D\rangle$ in two-dimensional
%
%
  Hilbert space $\cal {H}_D$. Its basis is  constituted
by   $|D_{1,2}\rangle$ eigenstates of $Q$
'pointer' observable with eigenvalues $q_{1,2}$.
The initial $D$ state is:            
$|D_0\rangle=\frac{|D_1\rangle+|D_2\rangle}{\sqrt {2}}$.
  $S$, $D$  interaction $\hat{H}_{S,D}$ starts at $t_0$
and finishes effectively at some  $t_1$;
for   Zurek Hamiltonian $H_{S,D}$ with suitable parameters$\cite {Z10}$
  it would result in  $S,\, D$  entangled final  state:
\begin {equation}
   \Psi_{S,D} = \sum a_i|s_i\rangle|D_i\rangle 
                                   \label {AA2}
\end {equation}
 in $O$ RF, it follows that
$ \bar{Q}=|a_1|^2-|a_2|^2$.  
 The measurement of
 $S$ eigenstate $|s_{1,2}\rangle$ results in factorized state
$\Psi^{1,2}=|s_{1,2}\rangle |D_{1,2}\rangle$ in which $Q$ has the eigenvalue
$q_{1,2}$.
 If $a_{1,2}\ne 0$, then $D$ is also described by the quantum
state $R_D$, but due to $S, D$ entanglement,
 it can't be completely factorized from $S$ state, so it's instructive
to use in the calculations $\Psi_{S,D}$ in place of $R_D$.
In our model   $D$, $O$ interaction starts at $t>t_1$
 and finishes at some $t_2$, during this time interval   
the information about $D$ state is transferred to $O$.
 Here we  suppose that $O$ can acquire all essential $D$ information
copiously, this assumption will be proved in sect. 3.


Let's remind how the measurement process  is described in
Information Theory \cite {G13}. The signal induced by the measured state and
 registrated by $O$ in event $n$ 
 is characterized by the array of real parameters called
 information pattern (IP) 
$J(n)=\{e_1,... ,e_l\}$.
The set of all possible $J$ constitutes the independent
 'information space',
which define $O$ recognition of measured states $\cite {S7}$.
In quantum case, some IP parameters, in principle, can be uncertain,
but this feature will be shown to be unimportant for our problem. 
 Consider, as the example, the  measurement of
 $S$ eigenstate $|s_{1,2}\rangle$.  
 $O$ supposedly percepts $|D_{1,2}\rangle$ state in event $n$ as IP:
$J^D_{1,2}=q_{1,2}$. 
 $Q$ eigenvalues $q_{1,2}$ are $D$ real properties$\cite {B1}$,
they correspond to the orthogonal projectors $P^D_{1,2}$.
 Hence  for $O$ the  difference between this $D$ states is the objective
or Boolean Difference (BD) $\cite {J3}$. It 
 is equivalent to the distinction between  the logical operands $Yes/No$,
 or that's the same   
between the values $1/0$ of some parameter $L_g$.

 For example, if the parameter $L_g=1/0$  for  $|D_{1,2}\rangle$,
then it corresponds to projector $P^D_1$.

     Now let's regard the measurement
 in case when $a_{1,2} \ne 0$,  i.e. $\psi_s$ is
 $|s_i\rangle$ superposition.
In this case, $\Psi_{S,D}$ state is  different from
$\Psi^{1,2}$ but it doesn't mean automatically that $O$ can discriminate
them as the different signals, this question demands the careful analysis.
Remind that the standard or  'Pedestrian' interpretation$\cite {Es2,J3}$ (PI)
 of QM claims that
  without the inclusion of Reduction Postulate into QM formalism, 
 $O$ would percept  $S$, $D$ entangled state $\Psi_{S,D}$ as 
  the  superposition of $|D_{1,2}\rangle$ states and so can discriminate them
from any of $|D_{1,2}\rangle$. 
 It supposed sometimes that it corresponds
the simultaneous observation of $J^D_{1,2}$,
 more realistically, one can expect  that  at least
   $\Psi_{S,D}$ is percepted by $O$ as some IP $J^s$,
 which should differ
 from   $J^D_{1,2}$.
 This assumption is the essence of famous
 'Schroedinger Cat' Paradox $\cite {J3}$.

However,
no such exotic outcomes are registrated experimentally,
in place of it, for the regarded kind of measurements
 $J^D_{1,2}$ are observed at random,
from that  it
is usually concluded that  Reduction Postulate should be added
 to QM formalism.
 Yet the situation isn't so simple
 and doesn't favor such prompt jump  to the  conclusions. 
Really, given PI implications are correct, $O$ should distinguish
 in a single event 
$\Psi_{S,D}$
 from both $|D_{1,2}\rangle$ states. 
Hence the relation of corresponding $O$ IPs should be characterized
 by BD, i. e. $J^s \ne J^D_{1,2}$. 
 So it should  be  at least  one  $D$ parameter (observable) $G^D$
  which value  $g_0$ for $\Psi_{S,D}$
is different from its values $g_{1,2}$ for $|D_{1,2}\rangle$.
 Roughly speaking,in this case it should be such $D$ parameter
$G_D$ which is equal to
$1/0$ depending on the presence/absence of $D_i$ superpositions. 
Meanwhile, in QM  all measurable parameters are related strictly to
 the observables   represented
by corresponding Hermitian Operators on $\cal H$ (or POV in general formalism).
Consequently, to verify the proposed PI hypothesis for
 $\Psi_{S,D}$ and $|D_{1,2}\rangle$,
 one should check the set (algebra) of
   $D$ PV observables $\{ G^D \}$ looking for the suitable candidates.
Yet  the simple analysis shows that there are no such  quantum $D$ observbles.
To demonstrate it,  suppose that such $G^O$ - Hermitian operator exists,
then it follows:
\begin {equation} 
   G^D\Psi_{S,D}=a_1|s_1\rangle G^D|D_1\rangle+a_2|s_2\rangle G^D |D_2\rangle
    =g_0\Psi_{S,D}
                      \label {E1}
\end {equation}
As easy to see, such equality fulfilled only for $G^D=I$. 
Any  $G^D$, which is sensitive to
the presence of superpositions,  
 corresponds to the   nonlinear operator on $\cal H_D$,
 so the observation of such difference seems to be incompatible
 with standard QM formalism. 
Consequently,  it seems  impossible for $O$ to
distinguish $|D_i\rangle$ from $R_D$ i.e. from $\Psi_{S,D}$ of (\ref {AA2})
 in a single event.  From that it's logical to conclude that $O$ would observe
one of $q_i$ outcomes in each event.
 For pure  $S$ ensemble it's reasonable to
assume that QM expectation value $\bar Q$ will be obtained by $O$
from the measurement of $S$  ensemble with the number of events
  $N \to \infty$. To fulfill such relation, $O$ should observe 
the stochastic $q_{1,2}$ outcomes with probabilities
 $\rm{P}_{1,2}=|\it a_{\rm 1,2}|^2$.
 This considerations put doubts
  on  the necessity of independent Reduction Postulate
in QM; the similar hypothesis was proposed by Wigner$\cite {W9}$. 
Note that the obtained results don't mean that $R_D$ is the probabilistic
 mixture of $|D_{1,2}\rangle$, rather $R_D$ can be characterized
 as their 'weak' superposition, stipulated by the entanglement of
 $S$, $D$ states. The possible role of joint $S, D$ observables will be
regarded  below,
but their account don't change the situation principally. 

\section { Measurements and Systems' Self-description } 

Now the information system  $O$ will be consistently
 considered as the quantum object.
In this case, MS is  described by the quantum state
 $\rho_{MS}$ relative to  some external  RF $O'$.
 In our model $O$ pure state is
 a vector in two-dimensional Hilbert space $\cal H_O$. Analogously to $D$,
we settle $O$ initial 
state $|O_0\rangle=\frac{|O_1\rangle+|O_2\rangle} {\sqrt 2}$ where
$|O_{1,2}\rangle$ are eigenstates of $O$ 'internal pointer' observable $Q_O$
with eigenvalues $q^O_{1,2}$. 
For suitable $D$, $O$ Hamiltonian $H_{D,O}$ one  obtains at  $t>t_2$: 
$$
   \Psi_{S,D,O} = \sum a_i|s_i\rangle|D_i\rangle |O_i\rangle
$$
As easy to see, $D$ states only double $S$ states for this set-up, so 
for the simplicity $D$ can be dropped from further considerations.
 In such scheme $S$ directly
interacts with $O$ by means of  Hamiltonian $H_{S,O}$, which result
 in the final state:
 \begin {equation}
   \Psi_{MS} = \sum a_i|s_i\rangle |O_i\rangle
                                   \label {A72}
\end {equation}
in  external RF $O'$.
Our aim  is to find the relation between this state and
 the information acquired by $O$.
Plainly,  the  measurement
  of arbitrary system  $S'$  by an information  system $O^I$
 can be considered as the  mapping of $S'$ states set $N_S$
 to  the set $N_O$ of $O^I$ internal states $\cite {M6}$.
 In Information Theory, this most general approach
is   described by the formalism of
systems' self-description called also 'measurement from inside' $\cite {S7}$.
 In its framwork,  $O^I$  considered  as
 the subsystem of larger system $\Xi_T=S', O^I$ with the states set $N_T$.
The information acquired by $O^I$  about $\Xi_T$ 
 (including $O^I$ itself) is
 described by $O^I$ internal state  $R_O$ called also
   $\Xi_T$ restricted state or restriction.
 For given $\Xi_T$ system $R_O$ is defined by 
the inference map $M_O$ of $\Xi_T$ state to $N_O$ set.
In quantum case $M_O$ derivation is the complicated problem for
any realistic $\Xi_T$ and now we shall turn to its detailed analysis.

 The important property of inference map $M_O$  
 is formulated by Breuer Theorem: if for two arbitrary $\Xi_T$
states $\Gamma,\Gamma'$
their restricted  states $R, R'$ coincide, then for $O^I$
 this $\Xi_T$
states are indistinguishable,  for any nontrivial $S',O^I$
at least one such pair of states exist $\cite {B5}$.
  In classical case, the origin of this  effect   is obvious:
  $O^I$ has less degrees of freedom  than $\Xi_T$ and hence
 can't discriminate all possible $\Xi_T$ states $\cite {S7}$.
 For  quantum systems
 $M_O$ ansatz should be derived from first QM principles,
 however, Schroedinger QM formalism only doesn't permit 
to derive
 $M_O(\Xi_T \to O^I)$ unambiguously and 
some additional inputs are needed for that.
For that purpose Breuer assumed phenomenologically
 that for  arbitrary  $\Xi_T$
its restricted  state  is equal to the
partial trace of $\Xi_T$ individual state over $S'$, i.e. is
$\Xi_T$ partial state on $O^I$.
In our MS set-up for  $\Psi_{MS}$  of (3) it gives:
\begin {equation}
   R^B_O=Tr_s  {\rho}_{MS}=\sum |a_i|^2|O_i\rangle\langle O_i|
      \label {AA4}
\end {equation}
Plainly, this ansatz excludes beforehand any
kind of stochastic $R_O$ behavior.
For MS mixed ensemble, induced by  the corresponding  $W^s$ gemenge,
the individual  MS states   differ from event to event:
\begin {equation}
\varsigma^{MS}(n)=|O_l\rangle \langle O_l|| s_l\rangle\langle s_l|
  \label {A44}
\end {equation}
where the frequencies of random $l(n)$ appearance in given event $n$
are stipulated by the
 probabilistic distribution $\rm P_{\it l}$ $=|a_l|^2$.
$O$ restricted   state for this mixed ensemble  is also stochastic:
in a given event 
$$
              R_O^{mix}(n)=.\xi^O_1.or.\xi^O_2.
$$
where $\xi^O_i =|O_i\rangle \langle O_i|$
 appears with the corresponding probability $\rm{P}_{\it i}$, so that
the ensemble of $O$ states described by the  gemenge
 $W^O_{mix}=\{\xi^O_i, \rm{P}_{\it i}  \}$.
 $R^{mix}_O(n)$ differs from $R^B_O$ in any event,
 hence   for the
restricted $O$ individual states the main condition of cited  theorem
is violated.  From that Breuer concluded
that  $O$ can discriminate the individual pure/mixed 
 MS states 'from inside',
so the collapse of pure state can't be observed by $O$ $\cite {B5}$.

Alternatively, we find that the information-theoretical considerations
  permit  to calculate
   MS restriction to $O$ directly and  unambiguosly; as will be shown,  
 the obtained results contradict to Breuer conclusion.
 To demonstrate it,  
  consider   the measurement of $S$ eigenstate $|s_{1,2}\rangle$,
it results in MS individual $\varsigma^{MS}$ state of (\ref {A44}),
 which restriction is $\xi^O_{1,2}$ with eigenvalues 
 $q^O_{1,2}$. Hence it's natural to conclude that  $O$ can identify 
 this states as IP  $   J^O_{1,2}=q^O_{1,2}$. The
 difference between
 $ \xi^O_i$ states is boolean (classical), because
 in QM formalism $\xi^O_i$ eigenvalues
$q^O_i$ are $O$  real properties \cite {B1}.
Now let's compare the  detection by $O$ of $\xi^O_I$ 
and $\Psi_{MS}$ of (\ref {A72}), i.e $R_O$.
 Note that the formal
 difference of two $O$ restricted states doesn't mean, in general,
 that this difference will be detected by $O$. Such  difference
is the necessary
but not sufficient condition for that, there should be also the specific
 $O$ observation $G^O$,
 which indicate this difference.
For  $R_O$ and $\xi^O_i$ the check of this hypothesis 
is analogous to
the approach described by  (\ref {E1}). 
Really, suppose that such $G^O$ exists, in QM framework, it should
be Hermitian PV  operator,
  from that $G^O$ should obey:
\begin {equation} 
   G^O\Psi_{MS}=a_1|s_1\rangle G^O|O_1\rangle+a_2|s_2\rangle G^O |O_2\rangle
    =g_0\Psi_{MS}
                      \label {EE22}
\end {equation}
Yet for $O$ observables such equality fulfilled only for $G^O=I$. 
 In the regarded  case, only the parameters corresponding
 to nonlinear operators can 
 establish BD between $R_O$ and $\xi^O_i$ states for $O$, but their
observability contradicts to standard QM axiomatic.
Consequently, $O$ can't distinguish  $R_O$ and $\xi^O_i$ states
and resulting MS restriction is equal to:
 \begin {equation}
         R_O=.\xi^O_1.or.\xi^O_2.   \label {B3}
\end {equation}
i.e. it coincides with $R_O^{mix}$ as the individual state.
 POV generalization of standard QM PV observables
doesn't change this conclusions.

It seems natural to expect that in the measurement of  $S$ pure ensemble $O$
should obtain ${Q}^l_O$ expectation values, which agree with
QM predictions for arbitrary $l$. To fulfill this condition,
 $O$ should observe the  collapse of pure MS state
to one of  $q^O_i$ at random with probability $\rm{P}_{\it i}$ $=|a_i|^2$, i.e.
the ensemble of $O$ states should be described by the  gemenge
 $W^O=\{\xi^O_i, \rm{P}_{\it i} \}$.
 It induces the corresponding $O$ IP ensemble 
$Z^O=\{ J^O_i, \rm {P}_{\it i} \}$ which describes
the collapse of $S$ pure state.  
Note however, that the assumption that
 the   probabilities of $q^O_i$ outcomes
follow QM ansatz, admitted here, isn't self-obvious \cite {H11}.
In general, it  should be proved in any new
theory of measurements \cite {B1}.
Leaving for the future the detailed proof for our theory,
    here we notice that for our MS, which consist of
 spin-$\frac{1}{2}$ objects only, such relations can be derived
from QM invariance relative to the space reflections and rotations.
 For example, for $a_1=a_2 e^{i\alpha}$
  the equality $\rm{P}_{\it 1}=\rm{P}_{\it 2}$ follows directly
 from QM reflection invariance. 

The difference between the pure and mixed $S$ states with the
same $\bar{S}_z$ is indicated by 
   'interference term' (IT) observables. 
 For our $S$ states they are
 $S_{x,y}$ linear forms.
  For example, if  $\frac{a_1}{a_2}$ is real, the maximal distinction  
reveals $S_x$ with
 $\bar{S}_x=\frac{1}{2} |a_1||a_2|$ for
 pure states and $\bar{S}_x=0$ for the mixture.
For MS entangled states such difference 
 can be revealed only by joint $S$,$O$ observables. As the example,
consider the symmetric IT for MS:
\begin {equation}
   B=|O_1\rangle \langle O_2||s_1\rangle \langle s_2|+j.c.
    \label {AA5}
\end {equation}
Being measured by external RF $O'$ via its interaction with $S$, $O$,
 it gives $\bar{B}=0$ for any  $|s_i\rangle$ incoming mixture,
 but  $\bar{B}\neq 0$  for  entangled MS states of (4).
For example, for   incoming  $S$ state $\psi^s_s$
 with $a_{1,2}=\frac{1}{\sqrt{2}}$, the resulting MS state
$\Psi^s_{MS}$ is $B$ eigenstate with eigenvalue $b_1=1$.
Hence in that case, $S_x$ is mapped to $B$. 
 However,  $B$ and any other IT   can't be directly measured by $O$
 'from inside', at least simultaneously with $S_z$, because they
 don't commute $\cite {B5}$.
Note also that the pure/mixed MS states with the same
$\bar{Q}_O$ can be discriminated even by external  $O'$
 only statistically, since the  corresponding  distributions of $B$ values (or
other ITs) overlap.  For example, for $\Psi^s_{MS}$ the probability 
 $\rm P_{\it B}(\it b_{1,2})=.5$
for $W^s$ mixture, so its $b$ distribution intersects largely
 with $\delta(b- b_1)$ distribution for $B$ eigenstate $\Psi^s_{MS}$. 
 Consequently, even $O'$   
 can discriminate  the pure/mixed MS states 
  only statistically for MS ensemble  with $N \to \infty$ but not
 in a single event.


It's well known that the decoherence of pure states by its environment $E$
is the important effect in quantum measurements $\cite {B1,Z10}$,
 we find yet that $O$ decoherence by $E$
doesn't play the principal role in our theory. However, 
its account stabilizes the described collapse mechanism additionally and  
defines unambiguously the preferred basis (PB)
 $\{\xi^O_i\}$ of $O$ 
final states used in our model.
Really, for the typical Hamiltonian of $O$,$E$ interaction
 $\cite {Z10}$,
 it follows that $\Psi_{MS}$ of (\ref {A72}) decoheres into
 MS,$E$ final state:
$$
    \Psi_{MS,E}=\sum a|s_i\rangle|O_i\rangle \prod\limits^{N_E}_{j}
           |E^j_i\rangle 
$$  
where $E^j$ are $E$ elements (atoms),
 $N_E$ is $E^j$ total number. If an arbitrary
$O$ pure state $\Psi_O$ is prepared, it will  also decohere in a very
 short time into the analogous $|O_i\rangle$
combinations, entangled with $E$. Hence, of all pure $O$ states,
only  $ \xi^O_i=|O_i\rangle$ are stable relative to $E$ decoherence.
Consequently, it advocates the choice of such $O$ states
as $O$ PB set, since in such environment $O$ simply can't
percept and memorize any other $O$ pure state during
 any sizable time interval.

\section {Conclusion}

Any consistent physical theory should not contain the logical
contradictions, it should be true also for the predictions of measurement
outcomes. Yet as was noticed by Wigner: 'The simultaneous observation
of two opposite outcomes of quantum experiments is nonsense' $\cite {W9}$.
It seems from our analysis that the structure of QM Observable Algebra 
by itself excludes such controversial observations even without the
 inclusion of Reduction Postulate into QM formalism.
 In addition, the formalism of systems'
self-description  permits to resolve  the old problem of
Heisenberg cut in quantum measurements, by the inclusion of the
information system into quantum formalism properly and on equal terms
with other MS elements.

 The most exciting and controversial 
question is whether this theory is applicable to the observations made by human
observer $O$, in particular, whether in this case IP $J^O$ describes
 the true $O$ 'impressions' about their outcomes ? This is open problem,
but at the microscopic level the human brain should obey QM laws
as any other object, so we don't see any serious reasons to make
the exceptions.
 Note that in our theory the brain
or any other processor $O$ plays only the passive role of signal receiver,
the real effect of information loss, essential for collapse,
occurs 'on the way', when the quantum signal passes
 through the information channel.

We conclude that standard Schr$\rm\ddot{o}$dinger QM formalism 
 together with the theory of systems' self-description
permit to obtain the 'subjective' collapse of  pure states
without implementation  of independent Reduction Postulate
into QM axiomatic.   
 In our approach the main
source of stochasticity  is  the principal
constraint on the transfer of  specific information in S$\,\to O$
 information channel.
  This  information
characterizes the purity of S state,
 because of its loss, $O$ can't discriminate the pure and mixed S states.
 As the result of this
information incompleteness,
the stochasticity of measurement outcomes appear,
 which can be interpreted as the analog of fundamental
 'white noise'.
 
The  interesting feature of this theory
 is that the same MS  state can be stochastic
in $O$ RF, but evolve linearly in $O'$ RF.
In particular, $\Psi_{MS}$ restriction to $O$ in $O$ RF is  stochastic 
state $R_O$ of (7), yet in $O'$ RF $O$ partial state is $R^B_O$ of (4),
i.e. is  the 'weak superposition'.
 The detailed explanation
of this effect is given by the formalism unitarily nonequivalent
 representations of  Algebraic QM $\cite {Em4}$. Here we notice only
that $O$ and $O'$ deal with different sets of MS observables, and so  
the transformation of MS  states between them is nonunitary.
 Obtained results agree well
with our calculations in $C^*$ Algebras formalism, in that approach
the inference map $M_O$ corresponds to the restriction of MS observable algebra
to $O$ (sub)algebra \cite {M8}.


\begin {thebibliography}{0}

\bibitem {B1}
   P.~Busch, P.~Lahti, P.~Mittelstaedt,
{ \it  Quantum Theory of Measurements}, Springer-Verlag, Berlin, 1996,
pp. 8--26

\bibitem {Es2}
  W.~D'Espagnat, {\it Found Phys.}  {\bf 20}, 1157--1169 (1990)

\bibitem {J3}
 J.~M.Jauch \it {Foundations of Quantum Mechanics},
Addison-Wesly, Reading, 1968, pp.  85--116

\bibitem {Em4}
  G.~Emch, \it {Algebraic Methods in Statistical Physics and
Quantum Mechanics}, Wiley, N-Y, 1972, pp. 71--89


\bibitem {B5}
  T.~Breuer,  \it {Synthese} {\bf {107}},   1--9 (1996)

\bibitem {M6}
 P.~Mittelstaedt, \it {Interpretation of
Quantum Mechanics and Quantum Measurement Problem},
Oxford Press, Oxford, 1998,  pp. 67--109

\bibitem {S7}
 K.Svozil, \it {Randomness and undecidability in Physics},
World Scientific, Singapour, 1993, pp. 46--87


\bibitem {M8}
        S.~Mayburov, \it {Int. J. Quant. Inf.} {\bf {5}}, 279--287 (2007); \,
Quant-ph/0506065


\bibitem {HHH}
 A.~S. Holevo, R.~F. Werner, \it {Phys. Rev.} {\bf  {A63}}, 
 032312--032326 (2000)
 
\bibitem {Z10} 
    W.~Zurek, \it {Phys. Rev.}  {\bf {D26}}, 1862--1876 (1982)

\bibitem {G13} U.Grenander, \it {Pattern Analysis}, 
Springer-Verlag, N-Y, 1978, pp 12--37

\bibitem {W9}
  E.~Wigner,  \it {Scientist speculates}, Heinemann, London, 1961, pp 47--59 

\bibitem {H11} 
     J.~B.Hartle, \it {Amer. J. Phys.} \bf  {36}, 704 (1968)
  


\end {thebibliography}
\end {document}